\documentclass[twocolumn, pre]{revtex4-1}

\usepackage{amssymb}
\usepackage{amsmath}
\usepackage{graphics}
\usepackage{graphicx}
\usepackage{mathrsfs}

\newcommand{\be}{\begin{equation}}
\newcommand{\ee}{\end{equation}}

\begin{document}

\title{The problem of shot selection in basketball}

\date{\today}
\author{Brian Skinner}
\affiliation{Fine Theoretical Physics Institute, University of Minnesota, Minneapolis, Minnesota 55455}

\begin{abstract}

In basketball, every time the offense produces a shot opportunity the player with the ball must decide whether the shot is worth taking.  In this paper, I explore the question of when a team should shoot and when they should pass up the shot by considering a simple theoretical model of the shot selection process, in which the quality of shot opportunities generated by the offense is assumed to fall randomly within a uniform distribution.  I derive an answer to the question ``how likely must the shot be to go in before the player should take it?", and show that this ``lower cutoff" for shot quality $f$ depends crucially on the number $n$ of shot opportunities remaining (say, before the shot clock expires), with larger $n$ demanding that only higher-quality shots should be taken.  The function $f(n)$ is also derived in the presence of a finite turnover rate and used to predict the shooting rate of an optimal-shooting team as a function of time.  This prediction is compared to observed shooting rates from the National Basketball Association (NBA), and the comparison suggests that NBA players tend to wait too long before shooting and undervalue the probability of committing a turnover.

\end{abstract} \maketitle

\section{Introduction}

In the game of basketball, the purpose of an offensive set is to generate a high-quality shot opportunity.  Thus, a successful play ends with some player from the offensive team being given the opportunity to take a reasonably high-percentage shot.  At this final moment of the play, the player with the ball must make a decision: should that player take the shot, or should s/he retain possession of the ball and wait for the team to arrive at a higher-percentage opportunity later on in the possession?

The answer to this question depends crucially on three factors: 1) the (perceived) probability that the shot will go in, 2) the distribution of shot quality that the offense is likely to generate in the future, and 3) the number of shot opportunities that the offense will have before it is forced to surrender the ball to the opposing team (say, because of an expired shot clock).  In this paper I examine the simplest model that accounts for all three of these factors.

Despite the game's lengthy history, the issue of shot selection in basketball has only recently begun to be considered as a theoretical problem \cite{Oliver2004bpr}.  Indeed, it is natural to describe the problem of shot selection as belonging to the class of ``optimal stopping problems", which are often the domain of finance and, more broadly, decision theory and game theory \cite{Moerbeke1976osa}.  A very recent work \cite{Goldman2011aad} has examined this problem using perspective of ``dynamic" and ``allocative" efficiency criteria.  The former criterion requires that every shot be taken only when its quality exceeds the expected point value of the remainder of the possession.  The second criteria stipulates that, at optimum, all players on a team should have equal offensive efficiency.  This allocative efficiency criterion is a source of some debate, as a recent paper \cite{Skinner2010poa} has suggested that the players' declining efficiency with increased usage implies an optimal shooting strategy that can violate the allocative efficiency criterion.  Further complications arise when considering ``underdog" situations, in which a team that is unlikely to win needs to maximize its chance of an unlikely upset rather than simply maximizing its average number of points scored per possession \cite{Skinner2011ssu}.  Nonetheless, Ref.\ \onlinecite{Goldman2011aad} demonstrates that players in the National Basketball Association (NBA) are excellent at shooting in a way that satisfies dynamic efficiency.  That is, players' shooting \emph{rates} seem to be consistent with their shooting \emph{accuracy} when viewed from the requirement of maximizing dynamic efficiency.  Still, there is no general theoretical formula for answering the question ``when should a shot be taken and when should it be passed up?".

Inspired by these recent discussions, in this paper I construct a simple model of the ``shoot or pass up the shot" decision and solve for the optimal probability of shooting at each shot opportunity.  This model assumes that for each shot opportunity generated by the offense the shot quality $p$ is a random variable, independent of all other shot opportunities, and is therefore described by some probability distribution.  For simplicity, following Ref.\ \onlinecite{Goldman2011aad}, all calculations in this paper assume that the probability distribution for $p$ is a flat distribution: that is, at each shot opportunity $p$ is chosen randomly between some minimum shot quality $f_1$ and some maximum $f_2$.  The best numerical definition for $p$ is the expected number of points that will be scored by the shot \footnote{
The possibility of offensive rebounds -- whereby the team retains possession of the ball after a missed shot -- is not considered explicitly in this paper, but one can think that this possibility is lumped into the expected value of a given shot.}; in other words, $p$ is the expected field goal percentage for a given shot multiplied by its potential point value (usually, 2 or 3).  If all shots are taken to be worth 1 point, for example, then $0 \leq f_1 \leq f_2 \leq 1$ .

The primary concern of this paper is calculating the optimal minimal value $f$ of the shot quality such that if players shoot if and only if the quality $p$ of the current shot satisfies $p > f$, then their team's expected score per possession will be maximized.  

I should first note that this ``lower cutoff" for shot quality $f$ must depend on the number of plays $n$ that are remaining in the possession.  For example, imagine that a team is running their offense without a shot clock, so that they can reset their offense as many times as they want (imagine further, for the time being, that there is no chance of the team turning the ball over).  In this case the team can afford to be extremely selective about which shots they take.  That is, their expected score per possession is optimized if they hold on to the ball until an opportunity presents itself for a shot that is essentially certain to go in.  On the other hand, if a team has time for only one or two shot opportunities in a possession, then there is a decent chance that the team will be forced into taking a relatively low-percentage shot.

So, intuitively, $f(n)$ must increase monotonically with $n$.  In the limit $n = 0$ (when the current opportunity is the last chance for the team to shoot), we must have $f(0) = f_1$: the team should be willing to take even the lowest quality shot.  Conversely, in the limit $n \rightarrow \infty$ (and, again, in the absence of turnovers), $f(n \rightarrow \infty) = f_2$: the team can afford to wait for the ``perfect" shot.  As I will show below, the solution for $f(n)$ at all intermediate values of $n$ constitutes a non-trivial sequence that can only be defined recursively.  I call this solution, $f(n)$, ``the shooter's sequence"; it is the main result of the present paper.

In the following section, I present the solution for $f(n)$ in the absence of turnovers.  Sec.\ \ref{sec:noclock} is concerned with calculating the optimal shot quality cutoff $f$ in ``pickup ball" situations, where there is no shot clock and therefore no natural definition of $n$, but there is a finite rate of turnovers.  Sec.\ \ref{sec:turn} combines the results of Secs.\ \ref{sec:f} and \ref{sec:noclock} to describe the case where there is a finite shot clock length as well as a finite turnover rate.  In Sec.\ \ref{sec:hazard} the sequence $f(n)$ is used to calculate the expected shooting rate as a function of time for an optimal-shooting team in a real game, where shot opportunities arise randomly over the course of the possession.  Finally, Sec.\ \ref{sec:data} compares these predicted optimal rates to real data taken from NBA games.  The comparison suggests that NBA players tend to wait too long before shooting, and that this undershooting can be explained in part as an undervaluation by the players of the probability of committing a turnover.

\section{The shooter's sequence} \label{sec:f}

In this section I calculate the optimal lower cutoff for shot quality, $f(n)$, for a situation where there is enough time remaining for exactly $n$ additional shot opportunities after the current one.  I also calculate the expected number of points per possession, $F(n)$, that results from following the optimal strategy defined by $f(n)$.  The effect of a finite probability of turning the ball over is considered in Secs.\ \ref{sec:noclock} and \ref{sec:turn}.

To begin, we can first consider the case where the team is facing its last possible shot opportunity ($n = 0$).  In this situation, the team should be willing to take the shot regardless of how poor it is, which implies $f(0) = f_1$.  The expected number of points that results from this shot is the average of $f_1$ and $f_2$ (the mean of the shot quality distribution): 
\be 
F(0) = \frac{f_1 + f_2}{2}
\ee

Now suppose that the team has enough time to reset their offense one time if they choose to pass up the shot; this is $n = 1$.  If the team decides to pass up the shot whenever its quality $p$ is below some value $y$, then their expected number of points in the possession is
\be 
F_y(1) = \frac{f_2 - y}{f_2 - f_1} \cdot \frac{y + f_2}{2} + \left( 1 - \frac{f_2 - y}{f_2 - f_1} \right) F(0).
\label{eq:optF1}
\ee
In Eq.\ (\ref{eq:optF1}), the expression $(f_2 - y)/(f_2 - f_1)$ corresponds to the probability that the team will take the shot, so that the first term on the right hand side corresponds to the expected points per possession from shooting and the second term corresponds to the expected points per possession from passing up the shot.  The optimal value of $p$, which by definition is equal to $f(1)$, can be found by taking the derivative of $F_y(1)$ and equating it to zero:
\be 
\left. \frac{d F_y(1)}{d y} \right|_{y = f(1)} = 0.
\label{eq:optf1}
\ee 

Combining Eqs.\ (\ref{eq:optF1}) and (\ref{eq:optf1}) gives $f(1) = F(0) = (f_1 + f_2)/2$.  In other words, the team should shoot the ball whenever the shot opportunity has a higher quality $p$ than the average of what they would get if they held the ball and waited for the next position.  This is an intuitive and straightforward result.  It can be extended to create a more general version of Eqs.\ (\ref{eq:optf1}) and (\ref{eq:optF1}).  Namely,
\be 
F_y(n) = \frac{f_2 - y}{f_2 - f_1} \cdot \frac{y + f_2}{2} + \left( 1 - \frac{f_2 - y}{f_2 - f_1} \right) F(n-1).
\label{eq:optF}
\ee
and 
\begin{eqnarray} 
\left. \frac{d F_y(n)}{d y} \right|_{y = f(n)} & = & 0.
\label{eq:optf}
\end{eqnarray}
Together, these two equations imply
\be 
f(n) = F(n-1).
\label{eq:Fequalsf}
\ee 
This is the general statement that a team should shoot the ball only when the quality of the current opportunity is greater than the expected value of retaining the ball and getting $n$ more shot opportunities.

The conclusion of Eq.\ (\ref{eq:Fequalsf}) allows one to rewrite Eq.\ (\ref{eq:optF}) as a recursive sequence for $f(n)$:
\be 
f(n+1) = \frac{[f(n)]^2 -2 f_1 f(n) + f_2^2}{2(f_2 - f_1)}.
\label{eq:ss}
\ee 
Along with the initial value $f(0) = f_1$, Eq.\ (\ref{eq:ss}) completely defines ``the shooter's sequence".  Surprisingly, considering the simplicity of the problem statement, this sequence $f(n)$ has no exact analytical solution.  Its first few terms and its asymptotic limit are as follows:
\begin{eqnarray}
f(0) &=& f_1 \nonumber \\
f(1) &=& (f_1 + f_2)/2 \nonumber \\
f(2) &=& (3 f_1 + 5 f_2)/8 \nonumber \\
f(3) &=& (39 f_1 + 89 f_2)/128 \nonumber \\
f(4) &=& (8463 f_1 + 24305 f_2)/32768 \nonumber \\
&...& \nonumber \\
f(n \rightarrow \infty) &=& f_2 \nonumber 
\end{eqnarray}
Note that in the limit where the team has infinite time, their shooting becomes maximally selective (only shots with ``perfect" quality $f_2$ should be taken) and maximally efficient (every possession scores $f_2$ points).

Since Eq.\ (\ref{eq:ss}) constitutes a recursive, quadratic map, it has no general solution \cite{Weissteinqm}.  Nonetheless, the expression for $f(n)$ can be simplified somewhat by writing it in the form
\be 
f(n) = \alpha(n) f_1 + \beta(n) f_2,
\ee 
where $\alpha(n)$ and $\beta(n)$ are separate recursive sequences defined by
\be 
\alpha(n+1) = \alpha(n) - \alpha(n)^2/2, \hspace{5mm} \alpha(0) = 1
\label{eq:alpha}
\ee
and
\be 
\beta(n) = \frac{1 + \beta(n-1)^2}{2}, \hspace{5mm} \beta(0) = 0,
\label{eq:beta}
\ee
respectively.  While $\alpha(n)$ and $\beta(n)$ have no analytical solution, in the limit of large $n$ they have the asymptotic behavior $\alpha(n) \simeq 2/n + \mathcal{O}(1/n^2)$ and $\beta(n) \simeq 1 - 2/n + \mathcal{O}(1/n^2)$.

\section{Optimal shooting without a shot clock} \label{sec:noclock}

In this section I consider ``pickup ball"-type situations, where there is no natural time limit to a possession.  In this case, the number of shot opportunities that the team can generate is limited only by their propensity to turn the ball over -- if the team attempts to continually reset the offense in search of a perfect shot they will eventually turn the ball over without taking any shots at all.

Thus, in these situations there is no natural definition of $n$, which implies that the solution for the optimal shot quality cutoff $f$ is a single number rather than a sequence.  Its value depends on the upper and lower values of the distribution, $f_1$ and $f_2$, and on the probability $p_t$ that the team will turn the ball over between two subsequent shot opportunities.  To calculate $f$, one can consider that the team's average number of points per possession, $F$, will be the same at the beginning of every offensive set, regardless of whether they have just chosen to pass up a shot.  The team's optimal strategy is to take a shot whenever that shot's quality exceeds $F$; \emph{i.e.}, $f = F$ as in Eq.\ (\ref{eq:Fequalsf}).  This leads to the expression
\be 
f = p_t \times 0 + (1-p_t) \left[ \frac{f_2 - f}{f_2 - f_1} \cdot \frac{f + f_2}{2} + \left( 1 - \frac{f_2 - f}{f_2 - f_1} \right) f \right].
\label{eq:optFnoclock}
\ee
In this equation, the term proportional to $p_t$ represents the expected points scored when the team turns the ball over (zero) and the term proportional to $1 - p_t$ represents the expected points scored when the team does not turn the ball over.  As in Eq.\ (\ref{eq:optF}), the two terms inside the bracket represent the points scored when the shot is taken and when the shot is passed up.

Eq.\ (\ref{eq:optFnoclock}) is a quadratic equation in $f$, and can therefore be solved directly to give the optimal lower cutoff for shot quality in situations with no shot clock.  This process gives
\be 
f = \frac{f_2 - f_1 p_t - \sqrt{p_t (f_2 - f_1)\left[2 f_2 - p_t(f_1 + f_2)\right]} }{1-p_t}.
\label{eq:fnoclock}
\ee
For $0 \leq p_t < 1$ and $0 \leq f_1 \leq f_2$, $f$ is real and positive.  In the limit $p_t \rightarrow 0$, Eq.\ (\ref{eq:fnoclock}) gives $f \rightarrow f_2$ (perfect efficiency), as expected.

\section{The shooter's sequence in the presence of turnovers} \label{sec:turn}

In this section I reconsider the problem of Sec.\ \ref{sec:f} including the effect of a finite turnover probability $p_t$.  This constitutes a straightforward generalization of Eqs.\ (\ref{eq:optF}) and (\ref{eq:optFnoclock}).  Namely,
\begin{eqnarray}
F(n) & = & (1-p_t) \times \left[ \frac{f_2 - f(n-1)}{f_2 - f_1} \cdot \frac{f(n-1) + f_2}{2} \right.  \nonumber \\
& & + \left. \left( 1 - \frac{f_2 - f(n-1)}{f_2 - f_1} \right) F(n-1) \right].
\label{eq:optFturn}
\end{eqnarray}
Simplifying this expression and using $f(n) = F(n-1)$ gives the recurrence relation
\be 
f(n) = (1-p_t) \frac{f(n-1)^2 - 2 f_1 f(n-1) + f_2^2}{2(f_2 - f_1)}.
\label{eq:fturn}
\ee
Together with the condition $f(0) = f_1$, Eq.\ (\ref{eq:fturn}) completely defines the sequence $f(n)$.

Unfortunately, the sequence $f(n)$ is unmanageable algebraically at all but very small $n$.  It can easily be evaluated numerically, however, if the values of $f_1$, $f_2$, and $p_t$ are known.  The first few terms of $f(n)$ and its limiting expression are as follows:
\begin{eqnarray}
f(0) &=& f_1 \nonumber \\
f(1) &=& (1-p_t)(f_1 + f_2)/2 \nonumber \\
f(2) &=& \frac{1-p_t}{8(f_2 - f_1)} \left\{ [5 - (2-p_t)p_t]f_2^2  \right. \nonumber \\
& &  \left. - 2 f_1 f_2 (1-p_t)^2 - f_2^2 (1-p_t)(3+p_t) \right\} \nonumber \\
&...& \nonumber \\
f(n \rightarrow \infty) &=& \frac{f_2 - f_1 p_t - \sqrt{p_t (f_2 - f_1)\left[2 f_2 - p_t(f_1 + f_2)\right]} }{1-p_t} \nonumber 
\end{eqnarray}
Notice that $f(n)$ approaches the result of Eq.\ (\ref{eq:fnoclock}) in the limit where many shot opportunities remain (\emph{i.e.} the very long shot clock limit).  

Overall, the sequence $f(n)$ has two salient features: 1) it increases monotonically with $n$ and ultimately approaches the ``no shot clock" limit of Sec.\ \ref{sec:noclock}, and 2) it generally calls for the team to accept lower-quality shots than they would in the absence of turnovers, since the team must now factor in the possibility that future attempts will produce turnovers rather than random-quality shot opportunities.

\section{Shooting rates of optimal shooters} \label{sec:hazard}

Secs.\ \ref{sec:f} -- \ref{sec:turn} give the optimal shot quality cutoff as a function of the number of shots remaining.  In this sense, the results presented above are useful for a team trying to answer the question ``when should I take a shot?".  However, these results do not directly provide a way of answering the question ``is my team shooting optimally?".  In other words, it is not immediately obvious how the shooter's sequence should manifest itself in shooting patterns during an actual game.

When analyzing the shooting of a team based on collected (play-by-play) data, it is often instructive to look at the team's ``shooting rate" $R(t)$.  The shooting rate (also sometimes called the ``hazard rate") is defined so that $R(t) dt$ is the probability that a team with the ball at time $t$ will shoot the ball during the interval of time $(t-dt, t)$.  Here, $t$ is defined as the time remaining on the shot clock, so that $t$ decreases as the possession goes on.  In this section I calculate the optimum shooting rate $R(t)$ implied by the results of Secs.\ \ref{sec:f} -- \ref{sec:turn}.  This calculation provides a means whereby one can evaluate how much a team's shooting pattern differs from the optimal one.

In order to calculate optimal shooting rate as a function of time, one should assume something about how frequently shot opportunities arise.  In this section I make the simplest natural assumption, namely that shot opportunities arise randomly with some uniform rate $1/\tau$.  For example, $\tau = 4$ seconds would imply that on average a team gets six shot opportunities during a 24-second shot clock.  I also assume that there is some uniform turnover rate $1/\tau_t$.  Under this set of assumptions, one can immediately write down the probability $P(t, n; \tau)$ that at a given instant $t$ the team will have enough time for exactly $n$ additional shot opportunities.  Specifically, $P(t,n;\tau)$ is given by the Poisson distribution:
\be 
P(n, t; \tau) = \left( \frac{t}{\tau} \right)^n \frac{e^{-t/\tau}}{n!}.
\label{eq:Poisson}
\ee 

The probability $p_t$ of a turnover between successive shot opportunities is given by
\be 
p_t = \int_0^\infty \left(1 - e^{-t'/\tau_t} \right) e^{-t'/\tau} \frac{dt'}{\tau} = \frac{\tau}{\tau_t + \tau}.
\label{eq:pt}
\ee
This integrand in Eq.\ (\ref{eq:pt}) contains the probability that there is at least one turnover during a time interval $t'$ multiplied by the probability that there are no shot attempts during the time $t'$ multiplied by the probability that a shot attempt arises during $(t', t'+dt')$, and this is integrated over all possible durations $t'$ between subsequent shot attempts.

In general, for a team deciding at a given time $t$ whether to shoot, the rate of shooting should depend on the proscribed optimal rate for when there are exactly $n$ opportunities left, multiplied by the probability $P(n, t; \tau)$ that there are in fact $n$ opportunities left, and summed over all possible $n$.  More specifically, consider that a team's optimal probability of taking a shot when there are exactly $n$ opportunities remaining is given by $[f_2 - f(n)]/(f_2 - f_1)$, where $f(n)$ is the shooter's sequence defined by Eq.\ (\ref{eq:fturn}).  The probability that the team should shoot during the interval $(t-dt, t)$ is therefore given by
\be 
R(t) dt = \frac{dt}{\tau} \sum_{n = 0}^{\infty} P(n, t; \tau) \frac{f_2 - f(n)}{f_2 - f_1}.
\ee 
Inserting Eq.\ (\ref{eq:Poisson}) gives
\be 
R(t) = \sum_{n = 0}^{\infty} \frac{t^n e^{-t/\tau}}{\tau^{n+1} n!} \cdot \frac{f_2 - f(n)}{f_2 - f_1}.
\label{eq:R}
\ee 
Since the sequence $f(n)$ has no analytical solution, there is no general closed-form expression for $R(t)$.  The corresponding expected average efficiency (points/possession) of a team following the optimal strategy is derived in the Appendix.

\begin{figure}[htb!]
\centering
\includegraphics[width=0.45 \textwidth]{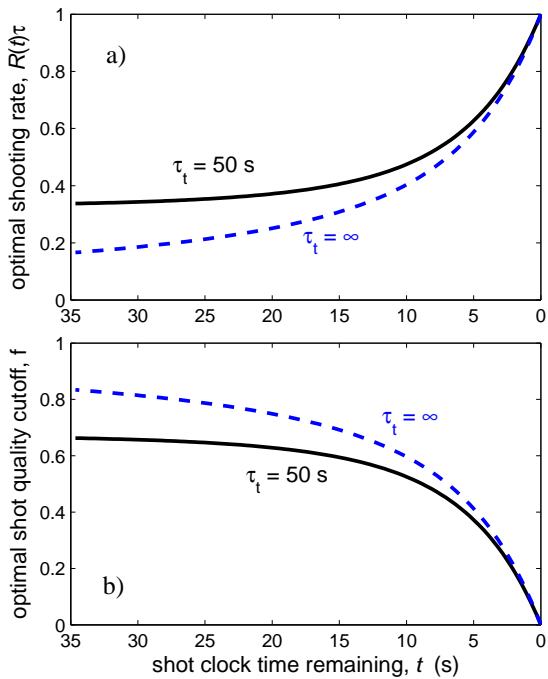}
\caption{a) Optimal shooting rate for a hypothetical team with $f_2 = 1$, $f_1 = 0$, $\tau = 4$ seconds, and $\tau_t = 50$ seconds, as given by Eq.\ (\ref{eq:R}).  The shooting rate $R(t)$ is plotted in the dimensionless form $R(t)\tau$, which can be thought of as the probability that a given shot that has arisen should be taken.  The dashed line shows the hypothetical shooting rate for the team in the absence of turnovers.
b) Optimal lower cutoff for shot quality, $f$, as a function of time for the same hypothetical team, both with and without a finite turnover rate.} \label{fig:hazardrates}
\end{figure}

As an example, consider a team that encounters shot opportunities with rate $1/\tau = 1/(4 \textrm{ seconds})$ and turns the ball over with rate $1/\tau_t = 1/(50 \textrm{ seconds})$.  Using the sequence defined in Eq.\ (\ref{eq:fturn}), one can evaluate numerically the shooting rate implied by Eq.\ (\ref{eq:R}).  This result is plotted as the black, solid line in Fig.\ \ref{fig:hazardrates}a, using $f_2 = 1$ and $f_1 = 0$.  In Fig.\ \ref{fig:hazardrates}a the optimal shooting rate is plotted as the dimensionless combination $R(t)\tau$, which can be thought of as the probability that a given shot should be taken if the opportunity arises at time $t$ (as opposed to $R(t)$, which is conditional on an opportunity presenting itself). For reference, I also plot the case where there are no turnovers, $\tau_t \rightarrow \infty$.  One can note that the finite turnover rate causes the optimal shooting rate to increase appreciably early in the shot clock.  In other words, when there is a nonzero chance of turning the ball over the team cannot afford to be as selective with their shots.

The optimal shooting rate can also be expressed in terms of the optimal lower cutoff for shot quality, $f$, as a function of time.  Since $R(t)\tau$ is the probability that a shot at $t$ should be taken, $f$ can be expressed simply as $f(t) = f_2 - R(t)\tau(f_2 - f_1)$.  This optimal lower cutoff is plotted in Fig.\ \ref{fig:hazardrates}b.  
A team that follows the optimal shooting strategy shown in Fig.\ \ref{fig:hazardrates} can be expected to score $0.64$ points per possession during games with a $24$-second shot clock [see Eq.\ (\ref{eq:Ft})], a significant enhancement from the value $0.5$ that might be naively expected by taking the average of the shot quality distribution.

In the limit of large time $t$ (or when there is no shot clock at all), as considered in Sec.\ \ref{sec:noclock}, the shooting rate $R(t)$ becomes independent of time and Eq.\ (\ref{eq:R}) has the following simple form:
\begin{eqnarray}
R & = & \frac{1}{\tau} \frac{f_2 - f}{f_2 - f_1}, \hspace{5mm} (\textrm{no shot clock}) \nonumber \\
 & = & \frac{1}{\tau_t} \left[ \sqrt{1 + \frac{2 f_2}{f_2 - f_1} \frac{\tau_t}{\tau} } - 1 \right].
\label{eq:Rnoclock}
\end{eqnarray}
Notice that when turnovers are very rare, $\tau_t \rightarrow \infty$, the shooting rate goes to zero, since the team can afford to be extremely selective about their shots.  

Eq.\ (\ref{eq:Rnoclock}) also implies an intriguingly weak dependence of the shooting rate on the average time $\tau$ between shot opportunities.  Imagine, for example, two teams, A and B, that both turn the ball over every $50$ seconds of possession and both have shot distributions characterized by $f_2 = 1$, $f_1 = 0$.  Suppose, however, that team A has much faster ball movement, so that team A arrives at a shot opportunity every 4 seconds while team B arrives at a shot opportunity only every 8 seconds.  One might expect, then, that  in the absence of a shot clock team A should have a shooting rate that is twice as large as that of team B.  Eq.\ (\ref{eq:Rnoclock}), however, suggests that this is not the case.  Rather, team B should shoot on average every $19$ seconds and the twice-faster team A should shoot every $12$ seconds.  The net result of this optimal strategy, by Eqs.\ (\ref{eq:fnoclock}) and (\ref{eq:pt}), is that team A scores $0.67$ points per possession while team B scores $0.57$ points per possession.  In other words, team A's twice-faster playing style buys them not a twice-higher shooting rate, but rather an improved ability to be selective about which shots they take, and therefore an improved offensive efficiency.

\section{Comparison to NBA data} \label{sec:data}

Given the results of the previous section, one can examine the in-game shooting statistics of basketball players and evaluate the extent to which the players' shooting patterns correspond to the ideal optimum strategy.  In this section I examine data from NBA games and compare the measured shooting rates and shooting percentages of the league as a whole to the theoretical optimum rates developed in Secs.\ \ref{sec:f}--\ref{sec:hazard}.  

The analysis of this section is based on play-by-play data from 4,720 NBA games during the 2006-2007 -- 2009-2010 seasons (available at http://www.basketballgeek.com).  Shots taken and points scored are sorted for all possessions by how much time remains on the shot clock at the time of the shot.  Following Ref.\ \onlinecite{Goldman2011aad}, possessions that occur within the last 24 seconds of a given quarter or within the last six minutes of a game are eliminated from the data set \footnote{
In such end-of-quarter or end-game situations, players are often not trying to optimize their offensive efficiency in a risk-neutral way.  An analysis of these ``underdog" situations is presented in Ref.\ \cite{Skinner2011ssu}.
}, along with any possessions for which the shot clock time cannot be accurately inferred.

\begin{figure}[htb!]
\centering
\includegraphics[width=0.45 \textwidth]{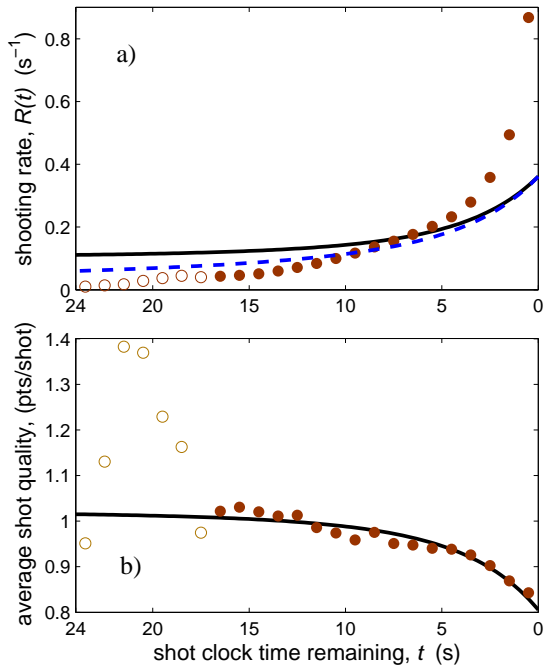}
\caption{A comparison between the theoretical optimum shooting strategy and data from NBA games.  a) The shooting rate as a function of shot clock time $t$.  The solid black line corresponds to the parameters $f_1 = 0.5$, $f_2 = 1.1$, $\tau = 2.8$ s, which are determined by a best fit to the shot quality data, using the NBA average turnover rate $\tau_t = 100.2$ seconds.  The dashed blue line corresponds to the same parameters except with the turnover rate $1/\tau_t$ set to zero.  b) The average shot quality (points per shot) as a function of $t$.  The solid line corresponds to the best fit curve to the filled symbols, from which the parameters for the solid black line in a) are determined.} \label{fig:data_compare}
\end{figure}

The resulting average shooting rate and shot quality (points scored per shot) are plotted as the symbols in Fig.\ \ref{fig:data_compare}a and b, respectively, as a function of time.  Open symbols correspond to shots taken during the first seven seconds of the shot clock, which generally correspond to ``fast break" plays during which the offense is not well-described by the theoretical model developed in this paper.  

In order to compare this data with the theoretical optimum behavior proscribed by the theories of Secs.\  \ref{sec:f}--\ref{sec:hazard}, one should determine the values $f_1$, $f_2$, $\tau$, and $\tau_t$ that best describe the average NBA offense.  This last parameter, the average time between turnovers, can be extracted directly the data: $\tau_t = 100.2$ seconds.  The other parameters can be determined only implicitly, by fitting the observed shooting rates and percentages to the theoretical model.

For the curves shown in Fig.\ \ref{fig:data_compare}, the following approach is employed.  First, the average shot quality for NBA teams is determined from the data as a function of time (Fig.\ \ref{fig:data_compare}b).  Then, the theoretical average shot quality $[f_2 + f(t)]/2$ of an optimal-shooting team is fit to this data in order to determine the best-fit values of $f_1$, $f_2$, and $\tau$, assuming optimal behavior.  This procedure gives $f_1 = 0.5$, $f_2 = 1.1$, and $\tau = 2.8$ seconds.  The corresponding fit line is shown as the solid curve in Fig.\ \ref{fig:data_compare}b.  The shooting rate $R(t)$ implied by these parameter values is then calculated and compared to the shooting rate measured from NBA games (Fig.\ \ref{fig:data_compare}a).  In this way one can compare whether the measured shooting \emph{rates} of NBA players are consistent with their shooting \emph{percentages}, within the assumptions of the theoretical model.

The result, as shown in Fig.\ \ref{fig:data_compare}a, is that NBA players seem noticeably more reluctant to shoot the ball during the early stages of the shot clock than is proscribed by the theoretical model.  With $15$ seconds remaining on the shot clock, for example, the average NBA team has a probability of only about $4\%$ of shooting the ball during the next second, whereas the optimal strategy suggests that this probability should be as high as $12\%$.  This observation is in qualitative agreement with the findings of Ref.\ \onlinecite{Goldman2011aad}, which concludes that under-shooting is far more common in the NBA than over-shooting.  As a consequence, NBA players are much more likely to delay shooting until the last few seconds of the shot clock, where they are likely to be rushed and their shooting percentages are noticeably lower. 

The price of this suboptimal behavior is reflected in the average efficiency $F$.  For NBA teams, the expected number of points per possession is $0.86$, or $0.83$ if one considers only possessions lasting past the first seven seconds of the shot clock.  In contrast, the optimal shooting strategy shown by the solid lines in Fig.\ \ref{fig:data_compare} produces $0.91$ points/possession for a $24$-second shot clock and $0.88$ points/possession for a 17-second clock (see the Appendix), even though it corresponds to the same distribution of shot quality.  This improvement of $0.05$ points/possession translates to roughly $4.5$ points per game.  According to the established ``Pythagorean" model of a team's winning percentage in the NBA \cite{JustinKubatko2007spa}, such an improvement can be expected to produce more than 10 additional wins for a team during an 82-game season.

One natural way to interpret the discrepancy between the observed and the theoretically optimal shooting behavior of NBA players is as a sign of overconfident behavior.  That is, NBA players may be unwilling to settle for only moderately high-quality shot opportunities early in the shot clock, believing that even better opportunities will arise later.  Part of the discrepancy can also be explained in terms of undervaluation of turnover rates.  If the players believe, for example, that they have essentially no chance of turning the ball over during the current possession, then they will be more likely to hold the ball and wait for a later opportunity.  This effect is illustrated by the dashed blue line in Fig.\ \ref{fig:data_compare}a, which shows the optimal shooting rate for the hypothetical case $\tau_t = \infty$ (the absence of turnovers).  This line is in significantly better agreement with the observed shooting rates at large $t$, which suggests that when NBA players make their shooting decisions early in the shot clock they do not account for the probability of future turnovers.

Of course, it is possible that much of the disagreement between the observed and theoretically optimum shooting rates can be attributed to an inaccuracy in the theory's assumption (in Sec.\ \ref{sec:hazard}) that shot opportunities arise randomly in time.  It is likely that NBA players often run their offense so as to produce more shot opportunities as the clock winds down.  This behavior would produce shooting rates that are weighted more heavily toward later times.  It is also likely that at very small time $t$ the theory's assumption of a uniform distribution of shot quality becomes invalid.  Indeed, in these ``buzzer-beating" situations the players' shots are often forced, and their quality is likely not chosen from the same random distribution as for shots much earlier in the shot clock.

In this sense, the theoretical result of Eq.\ (\ref{eq:R}) cannot be considered a very exact description of the shooting rates of NBA teams.  In order to improve the applicability of the model for real-game situations, one should account for the possibility of time dependence in the shot quality distribution ($f_1$ and $f_2$) and the rate of shot opportunities ($1/\tau$).  Such considerations are beyond the scope of the present work.  Nonetheless, the apparent sub-optimal behaviors illustrated in Fig.\ \ref{fig:data_compare} are instructive, and Eq.\ (\ref{eq:R}) may be helpful in determining how optimal strategy should adapt to changing features of the offense -- \textit{e.g.} an altered pace of play ($\tau$) or an improving/declining team shooting ability ($f_1$ and $f_2$) or a changing turnover rate ($\tau_t$).

If nothing else, the theories developed in this paper help to further the study of shot selection and optimal behavior in basketball, and may pave the way for a more complex theoretical model in the future.  In this way the problem of shot selection in basketball should be added to the interesting and growing literature on optimal stopping problems.  More broadly, the question of optimal behavior in sports provides an interesting, novel, and highly-applicable playground for mathematics and statistical mechanics.

\acknowledgments

I am grateful to M.\ R.\ Goldman and S.\ Redner for helpful discussions.

\appendix

\section{Expected offensive efficiency for a team following the optimal shooting rate}

In Sec.\ \ref{sec:hazard} the optimal shooting rate $R(t)$ was derived as a function of time [Eq.\ (\ref{eq:R})].  In this appendix I derive the number of points per possession that can be expected from a team with this optimal behavior.

For a shot taken at time $t$, the optimal lower cutoff for shot quality, $f(t)$, is given by $f(t) = f_2 - R(t)\tau(f_2 - f_1)$, as derived in Sec.\ \ref{sec:hazard}.  The corresponding average shot quality $\bar{p}(t)= [f_2 + f(t)]/2$ is given by
\be 
\bar{p}(t) = f_2 - \frac{f_2 - f_1}{2}R(t)\tau.
\ee  

To find the expected number of points per possession, one needs to know the probability that a shot will be taken during a given time interval $(t-dt, t)$.  This quantity can be written as $S(t; t_0)R(t)dt$, where $S(t; t_0)$ is the probability that the team still has the ball at time $t$ given that it gained possession at time $t_0$ (the beginning of the shot clock).

$S(t; t_0)$ can be derived by noting that the rate at which the current possession ends, $dS/dt$, is given by the sum of the shooting rate and the turnover rate multiplied by the probability that the possession has not ended already:
\be 
\frac{dS}{dt} = S(t;t_0) \left[ \frac{1}{\tau_t} + R(t) \right].
\ee
Rearranging this equation and integrating gives
\be 
S(t; t_0) = \exp\left[ - \int_t^{t_0} \left( R(t') + \frac{1}{\tau_t} \right) dt' \right].
\ee 
Given this expression for $S(t; t_0)$ one can calculate the expected number of points scored during the possession, $F$, by integrating the average shot quality at time $t$ multiplied by the probability of a shot being taken during $(t-dt, t)$ over all times $t$.  That is,
\be 
F = \int_0^{t_0} \bar{p}(t) S(t; t_0) R(t) dt .
\label{eq:Ft}
\ee
While a closed-form analytical expression for $F$ is not possible, Eq.\ (\ref{eq:Ft}) can easily be evaluated numerically.

\bibliography{overshooting}









\end{document}